\documentclass[5p,twocolumn]{elsarticle}

\usepackage{multirow}
\usepackage{gensymb}
\usepackage{amsmath}
\usepackage{amssymb}
\usepackage{booktabs}
\usepackage{comment}

\journal{Astroparticle Physics}

\begin{document}

\title{Phase-matching of multiple-cavity detectors for dark matter axion search}

\author[kaist]{Junu Jeong}
\author[ibs]{SungWoo Youn\corref{swyoun}}\ead{swyoun@ibs.re.kr}
\author[kaist]{Saebyeok Ahn}
\author[kaist]{Chanshin Kang}
\author[kaist,ibs]{Yannis K. Semertzids}
\address[kaist]{Department of Physics, Korea Advanced Institute of Science and Technology (KAIST), Daejeon 34141, Republic of Korea}
\address[ibs]{Center for Axion and Precision Physics Research, Institute for Basic Science, Daejeon 34047, Republic of Korea}
\cortext[swyoun]{Corresponding author.}

\date{\today}

\begin{abstract}
Conventional axion dark matter search experiments employ cylindrical microwave cavities immersed in a solenoidal magnetic field.
Exploring higher frequency regions requires smaller size cavities as the TM${_{010}}$ resonant frequencies scale inversely with cavity radius. 
One intuitive way to make efficient use of a given magnet volume, and thereby to increase the experimental sensitivity, is to bundle multiple cavities together and combine their individual outputs ensuring phase-matching of the coherent axion signal. 
We perform an extensive study for realistic design of a phase-matching mechanism for multiple-cavity systems and demonstrate its experimental feasibility using a double-cavity system.
\end{abstract}


\maketitle

\section{Introduction}

Axion, motivated by the R. Peccei and H. Quinn to solve to the CP problem in quantum chromodynamics of particle physics~\cite{bib:PQ}, is an attractive cold dark matter (CDM) candidate~\cite{bib:CDM}.
The current detection method, suggested by P. Sikivie, utilizes microwave resonant cavities placed in a strong magnet where axions are converted to radio-frequency (RF) photons~\cite{bib:Sikivie}.
The axion-to-photon conversion power is expressed as
\begin{equation}
P_{a\rightarrow\gamma\gamma} = g_{a\gamma\gamma}^2 \frac{\rho_a} {m_a}B_0^2VC\,{\rm{min}}(Q_L,Q_a)
\end{equation}
where $g_{a\gamma\gamma}$ is the axion-to-photon coupling, $\rho_a$ is the local halo density, $m_a$ is the axion mass, $B_0$ is the magnetic field, $V$ is the cavity volume, $C$ is the resonant mode form factor, and $Q_L$ and $Q_a$ are the loaded cavity and axion quality factors, respectively.
It is noted that the detection volume determined by the cavity size is an important experimental parameter for improving the experimental sensitivity.
An important quantity relevant to the experimental sensitivity is the signal-to-noise ratio (SNR), which is given by
\begin{equation}
{\rm{SNR}}\equiv \frac{P_{\rm{signal}}}{P_{\rm{noise}}} = \frac{P_{a\rightarrow\gamma\gamma}}{k_BT_{\rm{syst}}} \sqrt{\frac{t_{\rm{int}}}{\Delta f_a}},
\label{eq:snr}
\end{equation}
where $k_B$ is the Boltzmann constant, $T_{\rm{syst}}$ is the total system temperature, $t_{\rm{int}}$ is the integration time, and $\Delta f_a$ is the signal bandwidth.

Cavity-based axion search experiments typically employ a single resonant cavity fitting into a given magnet bore. 
However, exploring higher frequency regions requires smaller cavity sizes as the frequency of the resonant mode of our main interest, $\rm{TM_{010}}$, is inversely proportional to the cavity radius $R$, i.e., $f_{\rm{TM_{010}}}\sim R^{-1}$.
As conventional experiments rely on a single magnet with a fixed bore, an intuitive way to increase the detection volume for higher frequencies is to bundle multiple cavities together and combine the individual outputs coherently, which is referred to as ``phase-matching".
The idea of multiple-cavity design was introduced in 1990~\cite{bib:multi}; an experimental use of this design was attempted in 2000 using a quadruple-cavity detector~\cite{bib:multi_exp}, 
in which the design's methodological advantage was not fully addressed because reliability and increased complexity of operation were significant factors. 
Herein, we revisit this idea to verify that achieving phase-matching of an array of multiple cavities is realistically feasible and that the designed mechanism is certainly applicable to axion experiments in high frequency regions.
In this paper, we present an extensive study of a conceptual design for a phase-matching mechanism and demonstrate this design's experimental feasibility using a double-cavity detector.

\section{Receiver chain configuration}
There are three possible configurations in the design of a receiver chain for a multiple($N$)-cavity system. % within a single magnet.
One configuration comprises $N$ single-cavity experiments, consisting of $N$ independent complete receiver chains, in which the signals are statistically combined in the end, eventually resulting in a $\sqrt{N}$ improvement in sensitivity.
The other two configurations introduce a power combiner at an early stage of the receiver chain to build an $N$-cavity experiment, one with the first stage amplification taking place before the signal combination, the other with the signal combination preceding the first stage amplification.
Among the last two configurations, the former is characterized by $N$ amplifiers and a combiner, while the latter is characterized by a single amplifier and a combiner.
Table~\ref{tab:config} summarizes the schematic and characteristics of these three configurations.
In any design, an independent frequency tuning system is required for each cavity.
Assuming the axion signals from individual cavities are correlated, while the noises of system components are uncorrelated, configuration 2 gains an additional $\sqrt{N}$ improvement, yielding the highest sensitivity (see~\ref{app:snr}).
On the other hand, configuration 3 provides the simplest design with a sensitivity compatible with that of configuration 2~\cite{bib:combiner}.
As a simpler design is significantly beneficial especially for large cavity multiplicities, configuration 3 is chosen as the final design.

\begin{table}[h]
\caption{Possible configurations of the receiver chain for a multiple($N$)-cavity system. 
The cylinders, triangles, and D-shaped figures represent cavities, amplifiers, and combiners, respectively. 
$\rm{SNR_{sgl}}$ refers to the signal-to-noise ratio (SNR) of a single-cavity experiment.
The gain of the amplifiers is assumed to be sufficiently large.}
\begin{tabular*}{0.485\textwidth}{@{\extracolsep{\fill}}c@{}c@{}c@{}c}
\toprule
Configuration &1 & 2 & 3 \\
\midrule
\raisebox{1.5\height}{Schematic}
& \raisebox{-.07\height}{\includegraphics[width=0.1\textwidth, height=0.06\textwidth]{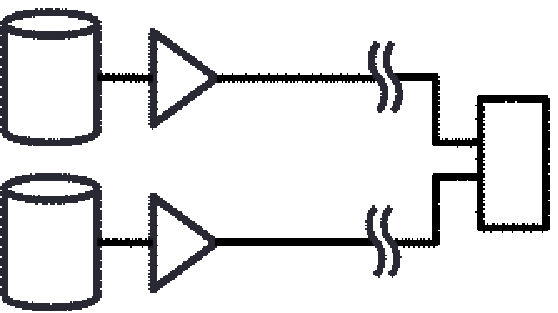}} 
& \raisebox{-.07\height}{\includegraphics[width=0.1\textwidth, height=0.06\textwidth]{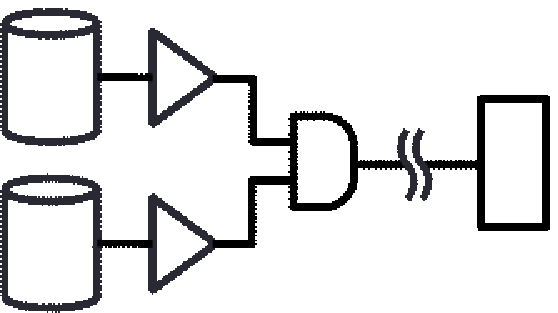}}
& \raisebox{-.07\height}{\includegraphics[width=0.1\textwidth, height=0.06\textwidth]{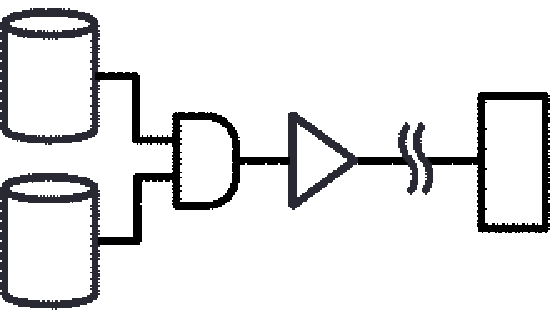}} \\
\multirow{2}{*}{Components} & $N$ complete & $N$ amplifiers & 1 amplifier \\
& chains & 1 combiner & 1 combiner \\ 
Sensitivity & $\sqrt N\cdot\rm{SNR_{sgl}}$ & $N\cdot \rm{SNR_{sgl}}$ & $\lesssim N\cdot\rm{SNR_{sgl}}$ \\
\multirow{2}{*}{Characteristic} & Individual & Hightest & Simplest \\
						& accessibility & sensitivity & design \\
\bottomrule
\end{tabular*}
\label{tab:config}
\end{table}

\section{Phase-matching}

\subsection{Frequency-matching}
Under the cosmological assumption that dark matter axions are virialized in our Galaxy, a condensate of CDM is represented by coherent oscillations of the axion field. 
The corresponding de Broglie wavelength of $10^1\sim10^3\,$m is much larger than the typical size of axion detectors ($<10^0$ m).
Individual detectors in a cavity array see approximately the same oscillation phase of the axion field and thus the electromagnetic (EM) fields of the cavity resonant modes coherently oscillate in a common external magnetic field.
This theoretical assumption enables a multiple-cavity approach to be experimentally plausible.

Since the axion mass (equivalently its frequency) is unknown, a cavity must be tuneable to scan the frequency range allowed by the cavity.
In addition, the cavities in an array must be independently tuned to the same frequency to enhance signal power, which is added linearly with the cavity multiplicity.
Failure in phase-matching broadens the width of the combined power spectrum, causing degradation of the effective quality factor, and eventually reduces the experimental sensitivity. 
This phase-matching in the frequency domain turns out to be the critical and challenging part of designing multiple-cavity systems.

The coherent axion signals extracted from individual cavities must interfere in a constructive manner at the combination level.
Phase-matching in the time domain, i.e., constructive interference, requires RF cables of the same length between the cavities and the power combiner.
Based on a pseudo-experiment, it is found that  in order to see more than 95\% of the ideally interfered signal, the cable length difference must be less than 18\,mm /$f\,\rm{[GHz]}$ for a quadruple cavity system.

Due to the large wavelength of the coherent axion field and the relatively facile achievement of constructive interference in signal combination, the phase-matching of a multiple-cavity system is in practice equivalent to frequency tuning of individual cavities to the same resonant frequency, which is referred to as frequency-matching.
The cavity frequency varies by changing the EM field distribution of the resonant mode, typically by means of a (pair of) dielectric or metal rod(s) inserted inside the cavity.
Unfortunately, ideal frequency-matching of multiple cavities is not possible mainly because of the machining tolerance for cavity fabrication and the non-zero step size of the tuning system.
Typical values of machining tolerance and step size of piezoelectric rotators are 50\,$\mu$m and 0.1\,m$\degree$, respectively.
These values correspond to a frequency difference of $\sim$10\,MHz and a frequency step of $\sim$0.5\,kHz for the TM$_{010}$ mode of a 5\,GHz resonant cavity.
Instead, a more realistic approach is to permit frequency mismatch up to a certain level at which the combined power is still sufficiently high that the resulting sensitivity is not significantly degraded.
We refer to this certain level as the frequency matching tolerance (FMT).

\subsection{Frequency matching tolerance}
\begin{figure}[h]
\centering
\includegraphics[width=0.36\textwidth]{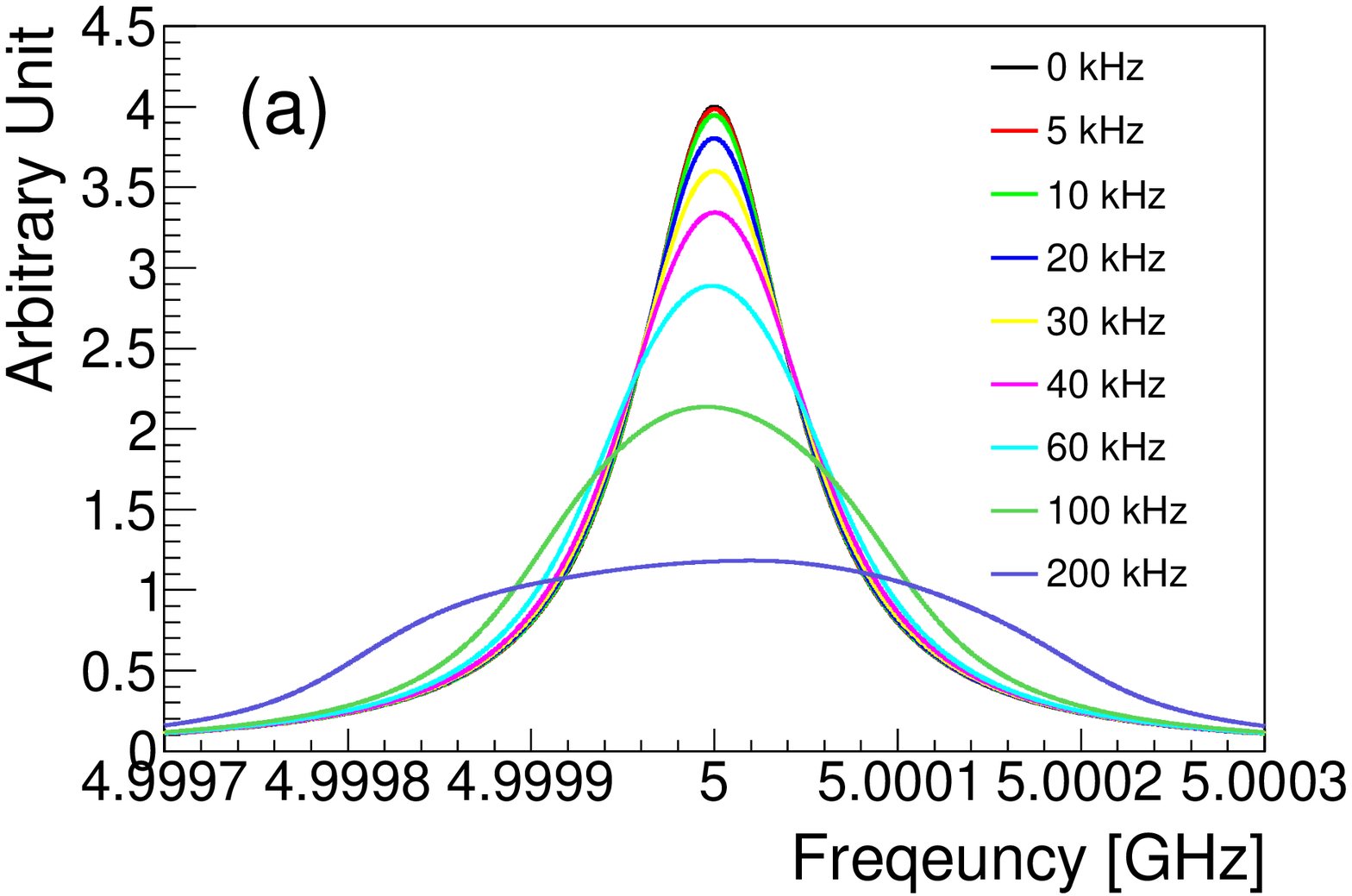}
\includegraphics[width=0.36\textwidth]{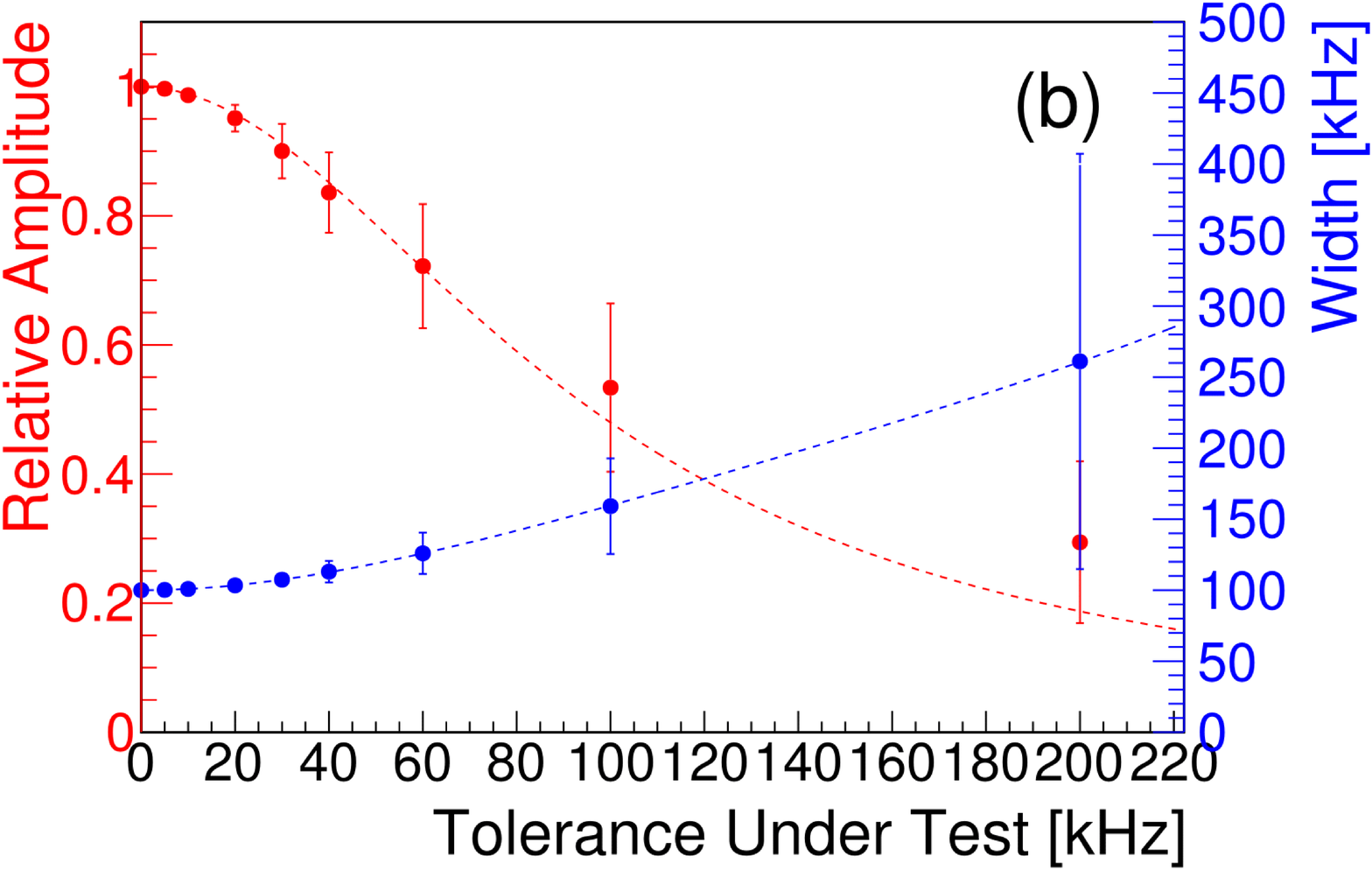}
\caption{(color online) (a) Combined power spectra averaged over 1,000 pseudo-experiments for several TUT values.
The power amplitude of each cavity is normalized to unity. 
(b) Relative power amplitude in red and full width at half maximum in blue as a function of TUT.
The error bars represent the statistical uncertainties.
These distributions are fitted with the Lorentzian and forth order polynomial functions, respectively.}
\label{fig:fmt}
\end{figure}

To determine FMT for multiple-cavity systems, a pseudo-experiment study is performed using a quadruple-cavity detector to search for a 5\,GHz axion signal.
The unloaded quality factors of the cavities are assumed to be the same at $Q_0=10^5$.
Supposing $Q_a\gg Q_0$, the signal power spectrum is expected to follow the Lorentzian distribution with its mean of 5\,GHz and half width of 50\,kHz.
Several values of frequency matching tolerance, or tolerance under test (TUT), are considered, i.e., 0, 5, 10, 20, 30, 40, 60, 100, and 200\,kHz, where 0\,kHz corresponds to ideal frequency-matching.
The cavities in the array are randomly tuned to the target frequency, 5\,GHz, following a uniform distribution with its center at 5\,GHz and half-width of the TUT under consideration.
Assuming each cavity is critically coupled, the individual power spectra are linearly summed up. 
The combined power spectrum is fitted with the Lorentzian function to obtain the amplitude and full-width at half maximum.
The procedure is repeated 1,000 times over which the combined power spectra are averaged.
Figure~\ref{fig:fmt} shows the distributions of the averaged combined power spectra for different TUT values (a) and displays the normalized amplitudes and full widths at half maximum as a function of TUT (b).
For a realistic approach, we impose the criteria that the relative amplitude of the combined power spectrum is greater than 0.95.
From this we find that the FMT is 21\,kHz for a system consisting of four identical cavities with $Q_0=10^5$ seeking for a 5\,GHz axion signal.
The FMT has a dependence on the cavity quality factor and target frequency; the dependence is generalized as FMT($Q_0, f$) = 0.42\,GHz/$Q_0\times f$\,[GHz].
It is noted that for $Q_0=10^6$ and $f=1$\,GHz, the FMT is 0.42\,kHz.
This tolerance is larger than the typical frequency step size of 0.1\,kHz for a 1\,GHz cavity, implying that frequency-matching for high $Q_0$ and low frequencies is still achievable.

\section{Tuning mechanism}
A multiple-cavity detector consists of $N$ cavities with identical dimensions and identical tuning systems.
The tuning system is comprised of a single dielectric rod inside a cavity and a single RF antenna through a hole on a cavity endcap.
One end of the tuning rod is attached to a piezoelectric rotational actuator so that the rod moves in a circle inside the cavity.
The resonant frequency of each cavity is tuned independently.
All the antennae, having the same dimensions, are sustained by a single holder which is attached to a piezoelectric linear actuator so that the coupling of the system is achieved in a global manner.
Global coupling is chosen at the sacrifice of a sensitivity loss of $<$0.5\%, which is estimated by taking into account a machining tolerance of 50\,$\mu$m and a variation in surface conductivity of a couple of percent.
These antennae are connected to a $N$-way power combiner by RF cables of the same length.
Finally, the output port of the combiner is connected to the first stage amplifier, beyond which the receiver chain is identical to that of a single-cavity experiment.

The basic principle of the tuning mechanism for a multiple-cavity system is the same as that for conventional single-cavity experiments.
The tuning mechanism relies on target frequency shift induced by rotating a single dielectric rod in parallel to the cavity axis, frequency matching performed by finely tuning the individual cavity frequencies, and critical coupling achieved by adjusting the depth of a single RF antenna in the cavity.
From an experimental point of view, the resonant frequency and the loaded cavity quality factor, $Q_L$, are determined by the reflection peak in the frequency spectrum of the scattering parameter ($S$-parameter) of a network analyzer.
Frequency-matching is assured by maintaining the minimum bandwidth of the reflection peak to give the maximum $Q$ value.
Critical coupling is characterized by the minimum reflection coefficient ($\Gamma$) in the scattering parameter and by the constant resistance circle passing through the center of the Smith chart~\cite{bib:smith}.
For a single-cavity system, $\Gamma$ is minimized when the system is critically coupled.
For a multiple-cavity system, on the other hand, the combined $\Gamma$ is minimized when the frequency-matching is successfully accomplished and the entire system (each cavity) is optimally coupled.
Therefore, the tuning mechanism for a multiple-cavity system consists of three steps: 
1) shifting the target frequency by simultaneously operating the rotational actuators; 
2) achieving frequency matching by finely manipulating the individual rotational actuators; and 
3) achieving critical coupling of the system by using the linear actuator to globally adjust the antenna depth.

\section{Experimental demonstration}
The feasibility of the tuning mechanism for multiple-cavity systems is experimentally demonstrated using a double-cavity detector at room temperature.
The detector is composed of two identical copper cavities with inner diameter of 38.8\,mm, whose corresponding resonant frequency is 5.92\,GHz, and unloaded quality factor of about 18,000. 
A single dielectric rod made of 95\% aluminium oxide (Al$_2$O$_3$) with a 4\,mm diameter is introduced to each cavity; a piezoelectric rotator is installed under the cavity to rotate the rod for frequency tuning.
With the tuning rod positioned at the center of the cavity, the resonant frequency decreases to 4.54\,GHz and $Q_L$ degrades to about 2,500 due to energy loss induced by the rod.
A pair of RF antennae, coupled to each cavity, are sustained by an aluminium holder attached to a linear piezoelectric actuator above the array of cavities.

The double-cavity system is assembled by connecting the two antennae to a two-way Wilkinson type power combiner that transmits signals to a network analyzer.
Using two sets of calibration kits, calibration is performed to compensate for the loss due to the antennae and cables.
Critical coupling of one cavity is achieved while the combiner input port, which the other cavity is connected to, is terminated with a 50\,$\Omega$ impedance terminator, and vice versa.
Two cavities are configured to be critically coupled at slightly different resonant frequencies. 
The initial values of the quality factor $Q_L$ and the $S$-parameter $S_{11}$ are measured.
After the system is re-assembled by connecting both the RF antennae to the combiner, 
the initial system is represented by two reflection peaks with $S_{11}=-6\,$dB in the $S$-parameter~\cite{bib:6dB} and two small circles with half-unit radius on the Smith chart.
At this stage we assume that the slight frequency difference is the consequence of the fabrication tolerance and the nonidentical frequency tuning systems.

\begin{figure}[h!]
\centering
\includegraphics[width=0.238\textwidth]{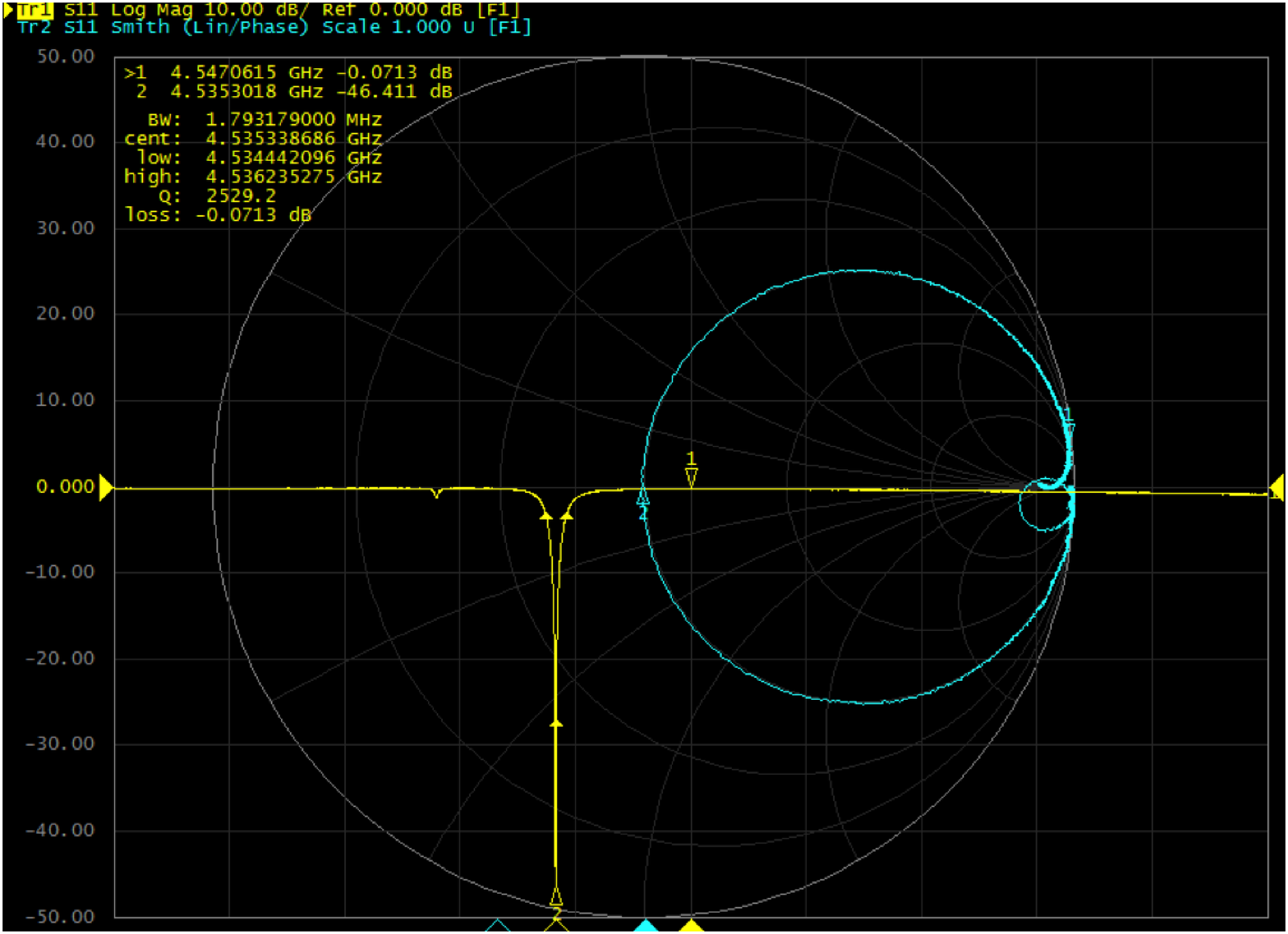}
\includegraphics[width=0.238\textwidth]{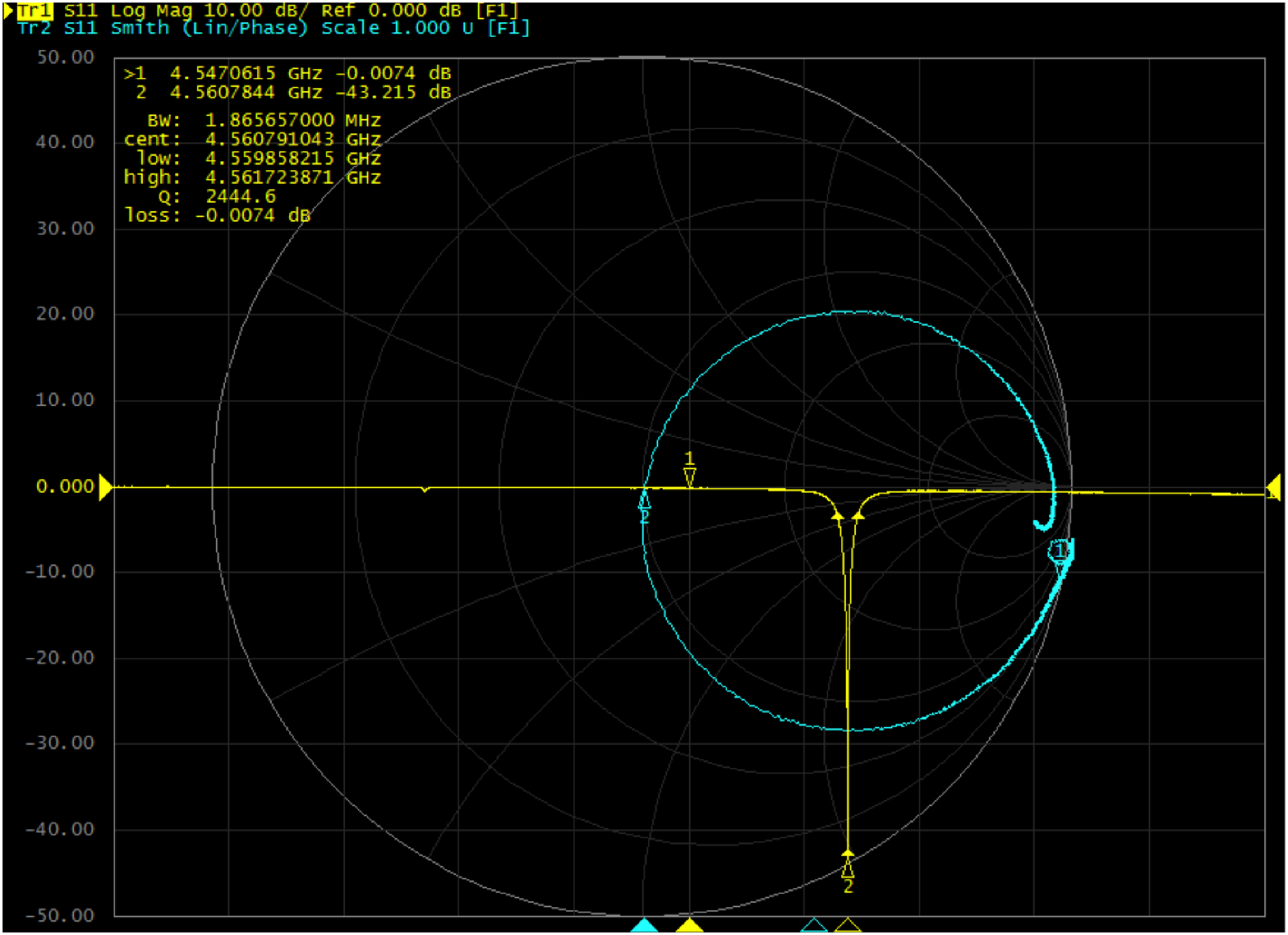}\\
\includegraphics[width=0.238\textwidth]{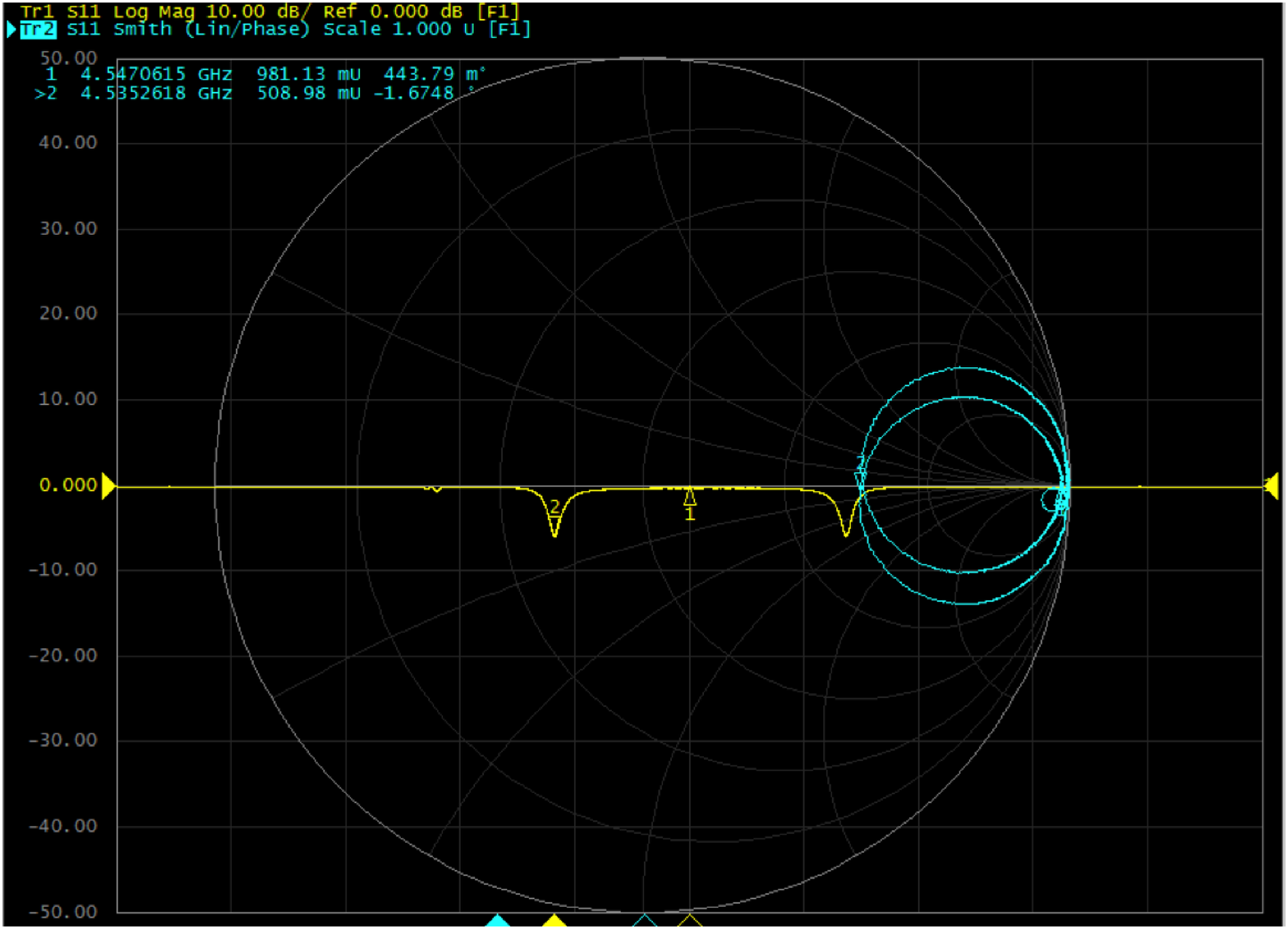}\\
\includegraphics[width=0.238\textwidth]{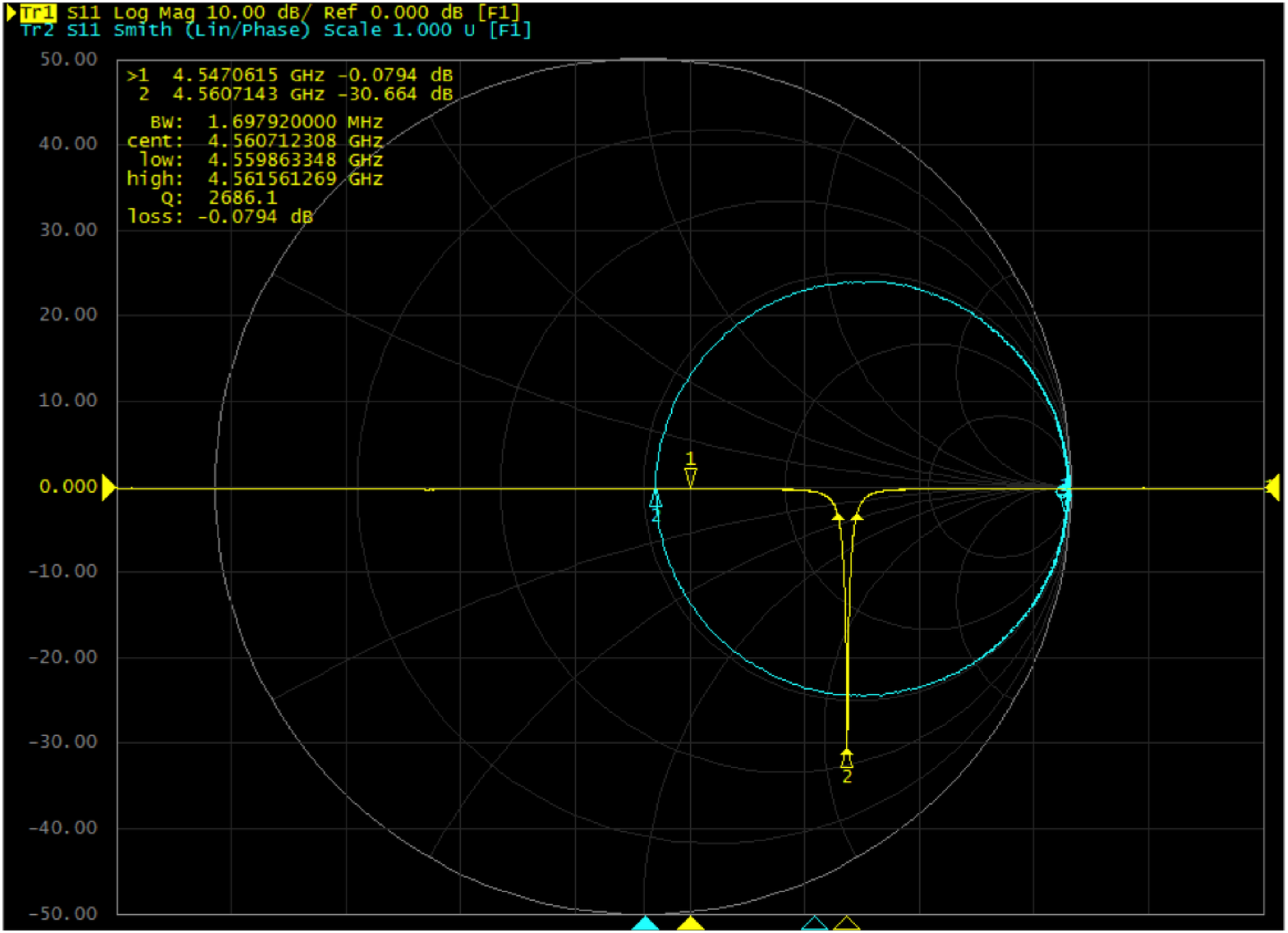}
\includegraphics[width=0.238\textwidth]{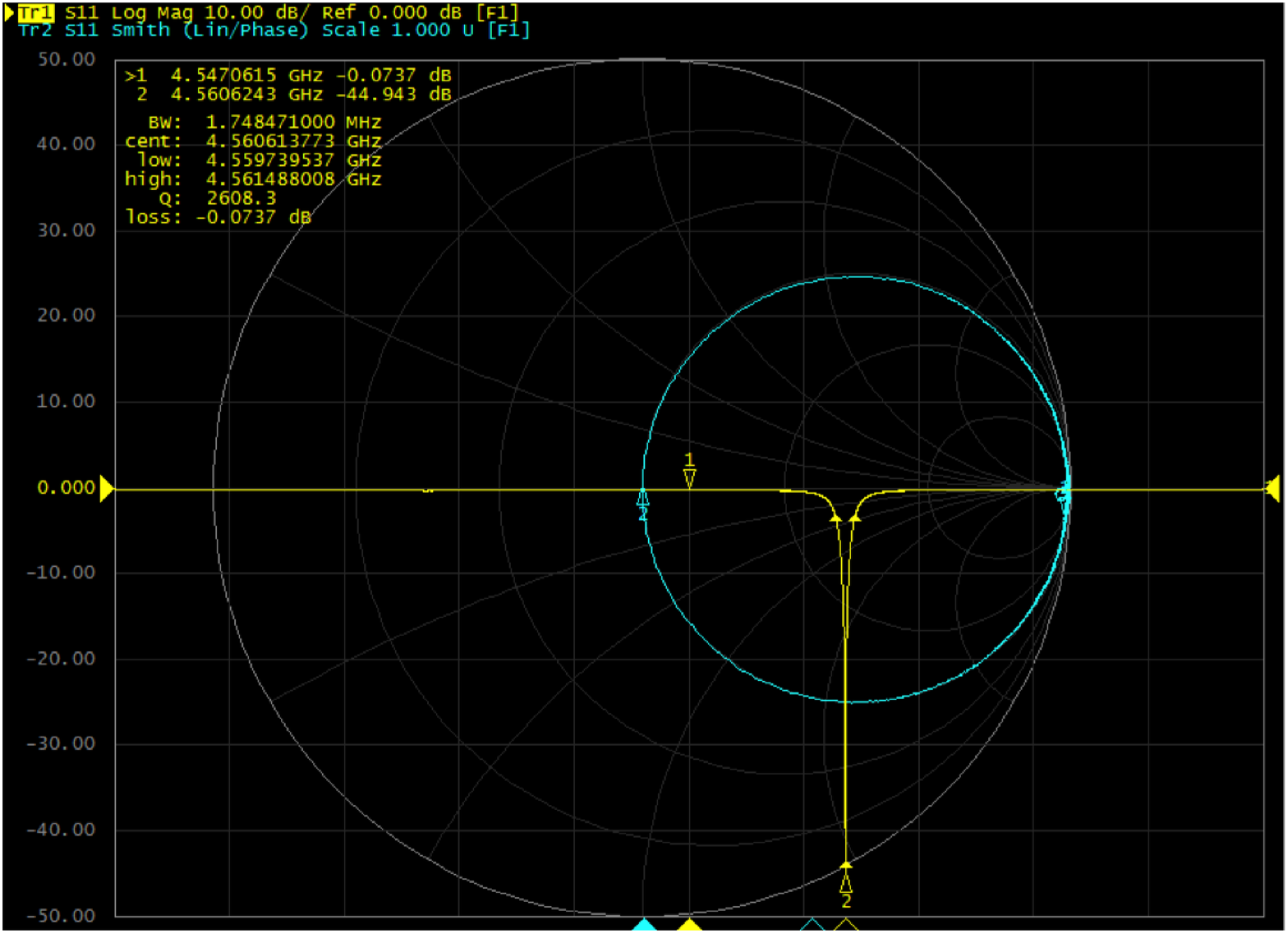}
\caption{(color online) Demonstration sequence of the tuning mechanism described in the text using a double-cavity system. 
Yellow solid lines are the scattering parameter, $S_{11}$, spectra in a logarithmic scale and cyan circles are their representations in the Smith chart. 
The first (second) cavity is critically coupled separately, and the initial values of $S_{11}$ and $Q_L$ are $-46.4$ $(-43.2)$\,dB and 2530 (2440) respectively (top left (right)). 
Two $-6$\,dB peaks and two small circles are observed after the system is fully assembled (middle). 
When the frequency is matched, a single deep peak and a single large circle are formed (bottom left).
Once critical coupling is accomplished in a global way, the peak becomes deeper and the circle passes through the center of the Smith chart. 
The final values of $S_{11}$ and $Q_L$ are $-44.4$\,dB and 2610 (bottom right).}
\label{fig:demo}
\end{figure}

In order to match the resonant frequencies, one of the rotational actuators is finely controlled until the combined reflection coefficient becomes minimized.
During this exercise, it is also observed that the two small circles on the Smith chart become a single larger circle.
Following this, the linear actuator is operated to adjust the antenna positions in a global manner to achieve critical coupling of the system until the reflection peak reaches its deepest point and the large circle passes through the center of the Smith chart. 
The final $Q_L$ and $S_{11}$ values are measured.
The consistency between the initial and final values convinces us that the tuning mechanism with frequency-matching is successful.
The demonstration sequence and characteristics of each step are shown in Fig.~\ref{fig:demo}. 
The few seconds of time consumption of the tuning mechanism also indicates that phase-matching for multiple-cavity systems is certainly feasible in real experiments.

\section{Conclusions}
We performed an extensive study of phase-matching for multiple-cavity systems as an effective way to increase the sensitivity of axion search experiments in high frequency regions. 
A receiver chain with signal combination preceding the first stage amplification is beneficial because of its simplest design and minimal degradation of signal power.
Frequency-matching among individual cavities is a key component of the phase-matching mechanism.
For a realistic approach, the frequency matching tolerance is introduced and numerically determined through peudo-experiments for a quadruple-cavity detector.
An experimental demonstration of the tuning mechanism (frequency matching and critical coupling), successfully conducted using a double-cavity detector, verifies its certain application to axion search experiments.

\begin{comment}
\section*{Supplementary Material}
See Supplemental Material at [URL will be inserted by publisher] for the video version of the demonstration of the phase-matching mechanism using the double-cavity system.
\end{comment}

\section*{Acknowledgments}
This work was supported by IBS-R017-D1-2017-a00 / IBS-R017-Y1-2017-a00.

\appendix
\section{Calculation of SNR} \label{app:snr}
For a single cavity system, denoting the axion signal by $S$, cavity noise by $N_c$, and gain and noise of the first stage amplifier by $G$ and $N_a$, respectively, the voltage output after preamplification is given by
\begin{equation}
V_{\rm{out}}=G(S+N_c)+N_a,
\end{equation}
and the SNR is obtained as
\begin{equation}
{\rm{SNR_{sgl}}}=\frac{P_{\rm{signal}}}{P_{\rm{noise}}}=\left(\frac{V_{\rm{signal}}}{V_{\rm{noise}}}\right)^2=\left(\frac{GS}{GN_c+N_a}\right)^2.
\end{equation}

For configuration 1, the SNR values of individual chains are the same as SNR$_{\rm{sgl}}$, and the total SNR becomes
\begin{equation}
\rm{SNR_1=\sqrt{\it{N}}\cdot SNR_{sgl}},
\end{equation}
with $\sqrt{N}$ representing the statistical combination after all.
For configuration 2, the voltage outputs at the intermediate stage, $V_{\rm{int}}$, between the peramplifiers and power combiner are the same as those of a single-cavity system.
Assuming the axion signals are correlated while the noises from the system components are uncorrelated, the voltage output after combination is calculated to be
\begin{equation}
V_{\rm{out}}=\frac{1}{\sqrt{N}}\left[NGS + \sqrt{N} (GN_c+N_a)\right],
\end{equation}
and the SNR becomes
\begin{equation}
{\rm{SNR_2}}=\left[\frac{NGS}{\sqrt{N}(GN_c+N_a)}\right]^2=N\cdot\rm{SNR_{sgl}}.
\end{equation}
For configuration 3, assuming no noise contribution from the combiner, the intermediate voltage output between combination and amplification is given by
\begin{equation}
V_{\rm{int}}=\frac{1}{\sqrt{N}}\left(NS + \sqrt{N}N_c\right),
\end{equation}
and the final voltage output is obtained as
\begin{equation}
V_{\rm{out}}=\frac{G}{\sqrt{N}}\left(NS + \sqrt{N}N_c\right) +N_a.
\end{equation}
This yields a SNR of
\begin{equation}
{\rm{SNR_3}}=\left[\frac{\sqrt{N}GS}{(GN_c+N_a)}\right]^2=N\cdot\rm{SNR_{sgl}}.
\end{equation}


\begin{thebibliography}{99}
\bibitem{bib:combiner} Configuration 3, where the combiner is placed prior to the first amplifier, degrades the SNR due to imperfection of the combiner. 
For instance, a system with a combiner with a noise figure of 0.5 and a amplifier with a gain of 12 and a noise figure of 6 yields a SNR reduction of $\sim$10\%.
\bibitem{bib:6dB} Two cavities are critically coupled at different resonant frequencies at which the combiner does not see the reflected signal. Note that for the reflected signal the combiner acts as a splitter so that the signal finally reflected through the common port is only one-fourth of the input signal.
\end{thebibliography}
\end{document}